\documentclass[journal]{IEEEtran}
\usepackage{amsmath,amsfonts}
\usepackage{algorithmic}
\usepackage{algorithm}
\usepackage{array}
\usepackage[caption=false,font=normalsize,labelfont=sf,textfont=sf]{subfig}
\usepackage{textcomp}
\usepackage{stfloats}
\usepackage{url}
\usepackage{verbatim}
\usepackage{graphicx}
\usepackage{cite}
\usepackage{booktabs}
\usepackage{xcolor}
\usepackage{soul}
\begin{document}

\title{Machine Learning Surrogate Model for Accelerating InGaAs/GaAs Quantum Well Laser Design}

\author{Paul~Verrinder,~\IEEEmembership{Member~IEEE,}, Jack~Markham
\thanks{P. Verrinder is with the Department of Electronics and Nanoengineering at Aalto University, Espoo, Finland. (email: paul.verrinder@aalto.fi).}
\thanks{J. Markham was with the Department of Electronics and Nanoengineering,
  Aalto University, Espoo, Finland. He is now with Université Jean Monnet, France.}
}

\markboth{}%
{Verrinder \MakeLowercase{\textit{et al.}}: Machine Learning Surrogate Model for InGaAs/GaAs Quantum Well Laser Simulation Acceleration}

\maketitle

\begin{abstract}
A machine-learning surrogate model based on a multi-layer perceptron neural network is used to predict six laser performance metrics for an indium gallium arsenide quantum well laser given epitaxial layer parameters as inputs. The surrogate model was trained on simulation data from a set of 4999 designs which were generated by varying epitaxial layer parameters and simulating each design with Harold to extract performance metrics. The model predicts all 6 chosen metrics with a coefficient of determination, $R^2$, greater than 0.91 on a held-out test set and demonstrates a $\sim 10^6 \times$ speed up compared to full device simulations. The model is used to search across 1 million potential epitaxial designs in the parameter space and find designs that minimize threshold current and internal loss. Using the surrogate, rather than full device simulation, reduces computation time for a million designs from 231 days to $\sim$2 minutes on a standard laptop.
It is also demonstrated that use of the Swish activation function, rather than ReLU, together with engineered input features - i.e. derived quantities calculated from the base model inputs - raise the mean $R^2$ from 0.88 to 0.94 without additional training data, and reduce the mean absolute percentage error from $1.78 \pm 0.08 \%$ to $1.17 \pm 0.09 \%$.
\end{abstract}


\section{Introduction}
\IEEEPARstart{S}{emiconductor} diode lasers are ubiquitous in a diverse range of applications and industries from optical communication and sensing \cite{arafinAdvancedInPPhotonic2018} to medical diagnostics, quantum sensing \cite{castroIntegratedModeHopFreeTunable2024}, and even high power lasers for fusion energy \cite{crumpDiodeLaserPumps2025b}. All diode lasers consist of a stack of epitaxially grown semiconductor layers to create a $pn$ junction with an active region embedded in the center. The most widely used design is the double heterostructure, where the active region is surrounded by higher bandgap confinement layers to provide both carrier and optical confinement \cite{coldrenDiodeLasersPhotonic2012}. III-V compound semiconductors such as gallium arsenide (GaAs), indium phosphide (InP) and gallium antimonide (GaSb), owing to their direct bandgap which makes them favorable for efficient light emission, are the most common substrates on which related binary, ternary or quaternary compounds can be grown for laser fabrication.

Laser diode epitaxial stacks can contain up to a dozen or more distinct layers. Every layer in the stack is defined by its material, composition (in the case of ternary and quaternary alloys), thickness, and doping concentration. In addition to other design factors such as laser cavity geometry, relevant laser performance metrics, such as threshold current or threshold modal gain, depend on the interplay between these epitaxial layers. 

Numerical solvers, such as Harold from Photon Design \cite{Harold}, are capable of simulating laser performance from a given epitaxy. However, full device simulations are slow enough that sweeping every parameter in the stack is impractical even for very coarse sweeps. Epitaxial design, then, usually starts with a `tried-and-true' layer stack and an iterative process is applied to optimize the design through targeted parameter sweeps and calculations - a process that relies on a combination of simulation and the engineer's intuition.

In this work, we demonstrate that a machine learning (ML) surrogate model can be trained to predict a set of selected laser performance metrics given epitaxial layers as input. The ML approach enables rapid, multi-variate optimization on the entire epitaxial layer stack since a single inference takes microseconds once the model is trained. Here the trained surrogate predicts threshold current $I_{th}$, threshold carrier density $N_{th}$, average internal loss $\alpha_i$, differential quantum efficiency $\eta_d$, peak wall plug efficiency $\eta_{wpe,~max}$, and active region temperature rise $\Delta T_{max}$. To train the model, we begin with a base epitaxy and define upper and lower limits for selected layer properties to define the epitaxial parameter space. The parameter space is then sparsely sampled within these bounds to generate a set of unique epitaxy designs which are simulated with Harold and become the dataset for training. The model learns the relationship between parameters, and can then be used to predict targets nearly instantaneously, enabling rapid exploration of the full design space for multi-objective optimization. 
ML and deep learning techniques have been applied to a variety of photonic design applications \cite{gentyMachineLearningApplications2021, maDeepLearningDesign2021}, and similar approaches have been used for optimization of quantum cascade laser (QCL) active regions \cite{correahernandezMachineLearningFramework2024, huEnhancingQuantumCascade2024, hernandezApplicationMachineLearning2023, hernandezGeneratingQuantumCascade2022} as well as gallium nitride (GaN) LED design \cite{jiangAdvancedDesignIIINitride2023}.
However, to our knowledge, this technique has not been applied to optimization of an entire epitaxial layer stack for QW edge emitting lasers. 

Here, we use a multi-layer perceptron (MLP) neural network (NN) to predict the six target metrics given doping profile, waveguide thickness and separate confinement heterostructure (SCH) layers for an indium gallium arsenide (InGaAs) quantum well (QW) laser. With a data set of 4999 samples (3499 for training and 750 each for validation and test), the model demonstrates good performance with a coefficient of determination, $R^2$, greater than 0.91 for all 6 targets when trained on 3499 input samples and a mean $R^2$ of 0.94 averaged across the 6 targets. We generated 5000 designs, however 1 simulation failed due to an unrelated software crash, so the final data set contains only 4999 designs. 

We also demonstrate that use of the Swish activation function \cite{ramachandranSearchingActivationFunctions2017} during training, rather than rectified linear unit (ReLU), and the inclusion of so-called ``engineered'' input features both improve model accuracy. 
Without Swish or engineered inputs, the mean $R^2$ remained at approximately 0.88; together they increase this to 0.94 without additional training data.
MLP training takes approximately 1 minute on a laptop with an Intel Core i5-13600H (12 cores / 16 threads, 4.8 GHz max turbo, 45 W) CPU using the MATLAB Deep Learning Toolbox \cite{MATLAB} \cite{DeepLearningToolbox}, and a single inference takes approximately 0.015 ms. This represents a speed-up of more than 6 orders of magnitude compared to a full device simulation with Harold which took an average of 20 seconds per design on the same laptop.

\section{Methods}

\subsection{Semiconductor Laser Design and Figures of Merit}
In this paper, we chose six values for the surrogate model to predict, all of which are useful quantities to know when designing or characterizing laser diodes \cite{coldrenDiodeLasersPhotonic2012}: threshold current $I_{th}$, threshold carrier density $N_{th}$, average internal loss $\alpha_i$, differential efficiency $\eta_d$, peak wall plug efficiency $\eta_{wpe\ max}$ and active region temperature rise $\Delta T_{max}$. 
We did not include lasing wavelength as a target because the epitaxy parameter space does not include variation in the active region which is the primary driver for wavelength. For the active region used here, the lasing wavelength is approximately 970 nm.
The average internal loss, $\alpha_i$, is a lumped parameter that contains all sources of optical loss within the laser cavity. Combined with mirror loss, $\alpha_m$, which can be calculated from the laser's cavity length and facet reflectivity, these quantities enable calculation of the threshold modal gain in the laser, $\Gamma g_{th}$,
\begin{equation}
\Gamma g_{th} = \alpha_i + \alpha_m
\end{equation}
where $\Gamma$ is the optical confinement factor in the QWs, $g_{th}$ is threshold gain, and $\alpha_m= \frac{1}{L} \ln(1/R),$ is the total mirror loss, where $L$ is the cavity length and $R$ is the total facet power reflectivity. In this work, we simulate Fabry-Perot (FP) lasers and the device geometry is held constant at 25 $\mu m$ wide with a 600 $\mu m$ long cavity and $R = 0.32$ to emulate standard cleaved facets. $\alpha_m$ is, therefore, a constant for all the simulated devices. Differential quantum efficiency, $\eta_d$, is obtained from the slope of the simulated LI curve and is also related to $\alpha_i$ and $\alpha_m$ through the following:
\begin{equation}
\eta_d =  \left[\frac{q}{h\nu} \right] \frac{dP_0}{dI} = \frac{\eta_i \alpha_m}{\alpha_i + \alpha_m},
\end{equation}
where $h\nu$ is photon energy, $q$ is the electron charge, $dP_0/dI$ is the slope of the LI curve (typically calculated just above threshold to avoid self-heating artifacts), and $\eta_i$ is internal quantum efficiency which can be calculated from $\alpha_i$ and $\alpha_m$ once $\eta_d$ is known.

\subsection{Epitaxy Parameter Space}
The base structure in this paper is a generic epitaxial layer stack for a GaAs-based laser with InGaAs QWs. This design consists of two 7 nm QWs in the active region, surrounded by 10 nm GaAs barriers, aluminum gallium arsenide (AlGaAs) SCH layers, and AlGaAs cladding layers. Additionally, a 50 nm indium gallium phosphide (InGaP) etch stop layer, lattice matched (LM) to GaAs, is placed directly below the active region. 
We chose to include this InGaP layer in the design as it facilitates selective active region removal and subsequent epitaxial regrowth for future active-passive integration \cite{koesterGaAsbasedPhotonicIntegrated2024, broxRealisationWidelyTuneable2017}. 
The simulated epitaxial layer stack is shown in Table \ref{tab:epitaxy}. To generate a training data set, some of these layer properties are varied; namely, the doping concentration in the cladding and SCH layers, thickness of the SCH and waveguide layers, and Al fraction in the SCH layers. These variables are indicated in bold in the table, and the bracketed quantities indicate the upper and lower limits of the range. Note that doping in the SCH layers is linearly graded and the two bracketed quantities indicate the range for the two endpoints of this linear grading. In total, there are 12 degrees of freedom (DoF) in the epitaxial parameter space. Note that the table shows the thickness of the waveguide layers, but in the data set generation, we actually set the total waveguide thickness (varying from 20-500 nm) and the fraction of the total waveguide thickness that sits on the p-side $wg_{np\ frac} = d_{pwg}/(d_{pwg}+d_{nwg})$.

\begin{table}[!t]
\caption{Simulated Epitaxial Layer Stack}
\label{tab:epitaxy}
\centering
\footnotesize
\setlength{\tabcolsep}{3pt}
\renewcommand{\arraystretch}{1.15}
\begin{tabular}{@{}llccc@{}}
\toprule
Layer & Material & $x$ & $d$ (nm) & $N$ ($10^{17}$) \\
\midrule
P-Cladding  & Al$_x$Ga$_{1-x}$As & 0.4                  & 1000               & \textbf{[1, 10]} \\
P-SCH       & Al$_x$Ga$_{1-x}$As & \textbf{[0.1, 0.35]} & \textbf{[10, 300]} & \textbf{[0.1, 5]}$\rightarrow$\textbf{[1, 10]} \\
P-Waveguide & GaAs               & --                   & \textbf{[20, 500]}$^{\mathrm{a}}$ & UID \\
Barrier     & GaAs               & --                   & 10                 & UID \\
QW          & In$_x$Ga$_{1-x}$As & 0.15                 & 7                  & UID \\
Barrier     & GaAs               & --                   & 10                 & UID \\
QW          & In$_x$Ga$_{1-x}$As & 0.15                 & 7                  & UID \\
Barrier     & GaAs               & --                   & 10                 & UID \\
Etch Stop   & InGaP              & LM                   & 50                 & UID \\
N-Waveguide & GaAs               & --                   & \textbf{[20, 500]}$^{\mathrm{a}}$ & UID \\
N-SCH       & Al$_x$Ga$_{1-x}$As & \textbf{[0.1, 0.35]} & \textbf{[10, 300]} & \textbf{[0.1, 5]}$\rightarrow$\textbf{[1, 10]} \\
N-Cladding  & Al$_x$Ga$_{1-x}$As & 0.4                  & 1000               & \textbf{[1, 10]} \\
\bottomrule
\end{tabular}
\vspace{2pt}
\begin{minipage}{\columnwidth}
\scriptsize
$^{\mathrm{a}}$Actual input parameters are total waveguide thickness (layer 3 + layer 10) and the fraction that sits on the p-side, $wg_{np\ frac}$, which is varied over [0.05, 0.95]. Bounds for total waveguide thickness: [20, 500] nm.
\end{minipage}
\end{table}

\begin{figure}[!t]
\centering
\includegraphics[width=3.25in]{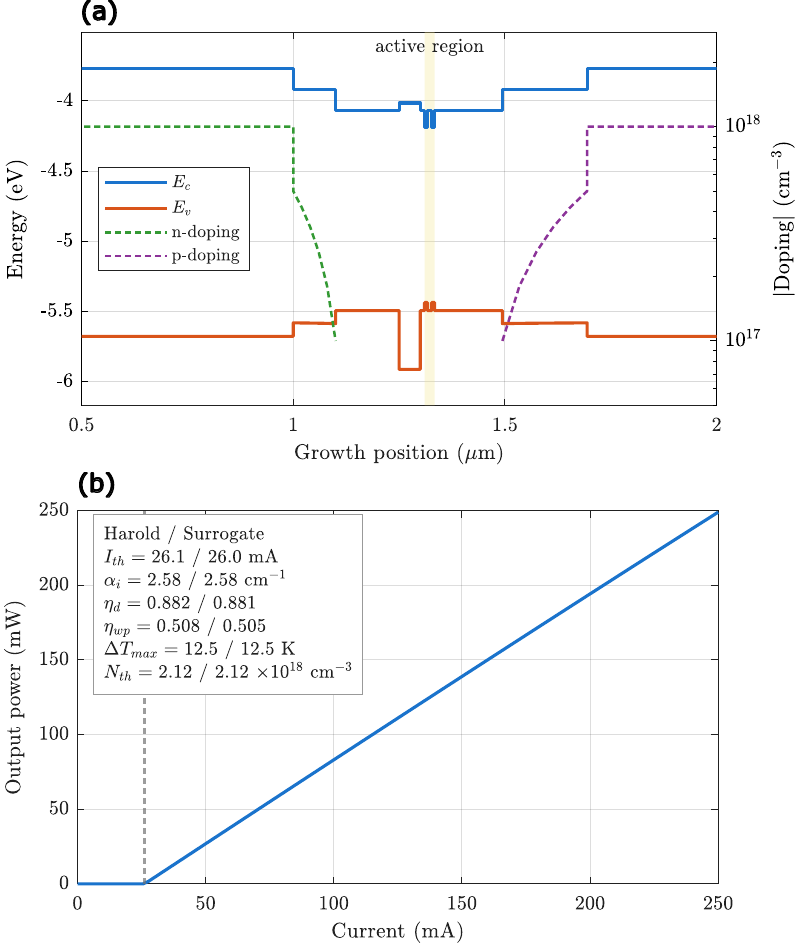}
\caption{(a) Band diagram for a representative design within the parameter
space and (b) LI curve simulated with Harold. Scalar target values from Harold and the surrogate model predicted values are shown in the text box.} 
\label{fig_bands}
\end{figure}

Fig. \ref{fig_bands}(a) shows the band diagram and doping profile for one possible design within this parameter space. Here, doping in both cladding layers is set to $1e18~cm^{-3}$ and in the SCH layers it is graded from $5e17~cm^{-3}$ at the SCH-cladding interface down to $1e17~cm^{-3}$ towards the center. The p-SCH thickness is $200~nm$, the n-SCH thickness is $100~nm$ and the Al fraction is set to 20\% in both. Total thickness of the waveguide layers is $300~nm$ and the QWs are placed in the center, just above the InGaP etch stop. Fig. \ref{fig_bands}(b) is the simulated light-current (LI) curve for this design. The inset text box shows the 6 scalar target values extracted from Harold, as well as the values predicted with the surrogate model, demonstrating near perfect agreement for this particular design. Accuracy of the surrogate model compared to Harold is discussed in detail in later sections. Notably, however, this exact design is not one of the training samples, so the comparison in Fig. \ref{fig_bands}(b) is an honest comparison for a design that the model has not seen during training.

\subsection{Dataset Sampling and Simulation}

The input data set was generated by assigning a value to each of the variable parameters in the layer stack, simulating that design in Harold, and extracting the 6 scalar targets from the simulation. This process was repeated for each design to generate a training data set consisting of 4999 input designs (i.e. 4999 unique layer stack definitions) and their associated Harold simulation outputs. The Harold software includes a python-based library for control, so the entire simulation and data extraction workflow was automated with python scripting.

\begin{figure}[!t]
\centering
\includegraphics[width=3in]{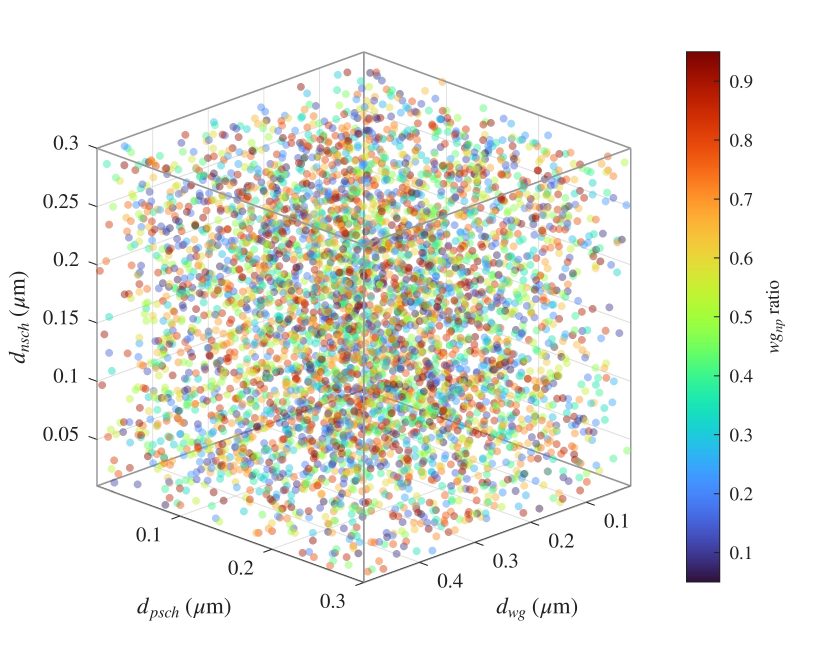}
\caption{Distribution of the LHS samples across $d_{wg}$, $d_{pSCH}$ and $d_{nSCH}$, colored by the waveguide p/n thickness}
\label{fig_lhs}
\end{figure}

Rather than pure random sampling within the variable ranges, we employed Latin Hypercube Sampling (LHS) to ensure even spacing between unique designs in the 12-dimensional space \cite{lohLatinHypercubeSampling1996}. LHS ensures that the generated samples are evenly spaced in the 12 sampled dimensions, which avoids unlucky clustering that could occur with purely random sampling, and ensures that designs near the edges are well represented. Fig. \ref{fig_lhs} shows how the samples are distributed for 3 of the dimensional parameters, $d_{wg}$, $d_{pSCH}$, and $d_{nSCH}$, with the points colored by $wg_{np\ frac}$.

As mentioned above, the commercial laser simulation software, Harold was used to generate the simulated data set. Harold is a heterostructure simulator for FP lasers which uses well-established physical models to account for a wide range of physical effects \cite{al-muhannaWaveguideDesignOptimization2011, paolettiWavelengthstabilizedDBRHighpower2020}. In this paper, the simulated laser geometry is fixed with a cavity length of 600 $\mu m$ and width of 25 $\mu m$. Simulations were run in self-heating mode with the current swept from 0 to 250 mA in each simulation and the stage temperature held at $25^\circ$ C. The epitaxial layer stack was based on Table \ref{tab:epitaxy} with the 12 variable parameters defined as variables in the Harold project file. Python scripting was used to automate the simulation loop for the 5000 input designs. This loop consists of the following steps: 1. set epitaxy variables $\rightarrow$ 2. run simulation $\rightarrow$ 3. extract the six scalar targets from the simulation results $\rightarrow$ 4. save input parameters (epi stack) and outputs (the 6 scalar targets) to a .csv formatted table $\rightarrow$ 5. repeat for the next design.

\subsection{Engineered Input Features}
The default, or ``raw'', inputs are the 12 variables from Table \ref{tab:epitaxy}. However, without adding more data, we can also include engineered input features that are derived from these 12 and which have some physical significance. These are displayed in Table \ref{tab:features}. In the definitions, $\bar{N}_p = \frac{1}{2}(\lvert N_{p1}\rvert + \lvert N_{p2}\rvert)$ and $\bar{N}_n = \frac{1}{2}(N_{n1} + N_{n2})$ are the graded SCH doping levels averaged across their two endpoints.

Internal loss is heavily influenced by free-carrier absorption (FCA) and is generally strongest in the SCH layers, which are doped but still interact with the optical mode \cite{bulashevichEffectFreecarrierAbsorption2007}. As a crude proxy for FCA in these layers, we have defined a sheet doping density for both the n- and p-SCH layers as  $N_{s,p\mathrm{SCH}} = \bar{N}_p \, d_{p\mathrm{SCH}}$ and $N_{s,n\mathrm{SCH}} = \bar{N}_n \, d_{n\mathrm{SCH}}$, as well as the sum of these two. We also define the asymmetry in the average SCH doping levels as $A_{\mathrm{SCH}}$. The cladding layers are in the evanescent tail so their contribution to FCA loss is very weak, although not technically zero. More importantly, however, the cladding layers influence diode series resistance which is directly related to the Ohmic self-heating that determines $\Delta~T_{max}$. 
In addition to the cladding doping defined in Table \ref{tab:epitaxy}, we also define the sum of the magnitudes of the doping concentration in the two cladding layers, $N_{\mathrm{clad},\mathrm{tot}}$. 
During simulation, we only pass the total waveguide thickness and the fraction that sits on the p-side; here we also define the n-side and p-side waveguide thicknesses, $d_{n\mathrm{wg}}$ and $d_{p\mathrm{wg}}$.
The values in Table \ref{tab:features} are clearly redundant with the 12 parameters in Table \ref{tab:epitaxy}, however, it costs almost nothing computationally to calculate them for each simulation and passing them as inputs prevents the model from needing to learn these relationships. The precise physical significance of these quantities is less relevant than the fact that they are relationships that the model will learn from the data regardless and adding them as inputs makes that learning free. As demonstrated in the next section, adding these during training results in measurable performance improvements.

\begin{table}[!t]
\caption{Engineered Features Derived from the 12 Sampled Design Variables}
\label{tab:features}
\centering
\footnotesize
\setlength{\tabcolsep}{3pt}
\renewcommand{\arraystretch}{1.2}
\begin{tabular}{@{}lll@{}}
\toprule
Feature & Description & Definition \\
\midrule
$N_{s,p\mathrm{SCH}}$ & p-SCH sheet doping         & $\bar{N}_p\, d_{p\mathrm{SCH}}$ \\
$N_{s,n\mathrm{SCH}}$ & n-SCH sheet doping         & $\bar{N}_n\, d_{n\mathrm{SCH}}$ \\
$N_{s,\mathrm{SCH}}$  & Total SCH sheet doping     & $N_{s,p\mathrm{SCH}} + N_{s,n\mathrm{SCH}}$ \\
$A_{\mathrm{SCH}}$       & SCH doping asymmetry       & $(\bar{N}_p - \bar{N}_n)/(\bar{N}_p + \bar{N}_n)$ \\
$N_{\mathrm{clad},\mathrm{tot}}$  & Total cladding doping    & $|N_{p\mathrm{clad}}| + |N_{n\mathrm{clad}}|$ \\
$d_{p\mathrm{wg}}$       & p-side waveguide thickness & $d_{\mathrm{wg}}\, wg_{np\ frac}$ \\
$d_{n\mathrm{wg}}$       & n-side waveguide thickness & $d_{\mathrm{wg}}(1 - wg_{np\ frac})$ \\
\bottomrule
\end{tabular}
\end{table}

\subsection{Model Architecture and Training}

With the target datasets constructed, ML surrogate models were developed to map the input epitaxial parameters $\mathbf{x}$ to the output laser property values $\mathbf{y}$. 
The architecture evaluated in this work is the MLP, a feedforward artificial neural network capable of learning complex, non-linear relationships, with either 12 or 19 inputs (depending on whether the 7 engineered features are used) in the input layer. 
The network consists of 3 hidden layers of 48 neurons wide, each followed by batch normalization and the activation function, and an output layer for the 6 scalar target metrics. The network was trained with the Adam optimizer at an initial learning rate of $10^{-3}$ and a mini-batch size of 128. Training ran for at most 400 epochs, with early stopping where the validation loss is evaluated once per epoch and training halts after 25 epochs without improvement; the weights from the epoch with the lowest validation loss were retained. In practice, training terminated after roughly 200 epochs. 
Inputs are z-scored at the input layer and targets are standardized to zero mean and unit variance on the training set.
The three hidden layer MLP architecture follows reference \cite{correahernandezMachineLearningFramework2024}; we experimented with hidden layer widths of 64 and 32 neurons, both of which performed worse than the 48 neuron width, but did not otherwise perform a thorough hyperparameter search. The MATLAB code for training is available at reference \cite{verrinder2026code} and contains the complete configuration for the model.

\begin{figure}[!t]
\centering
\includegraphics[width=3.25in]{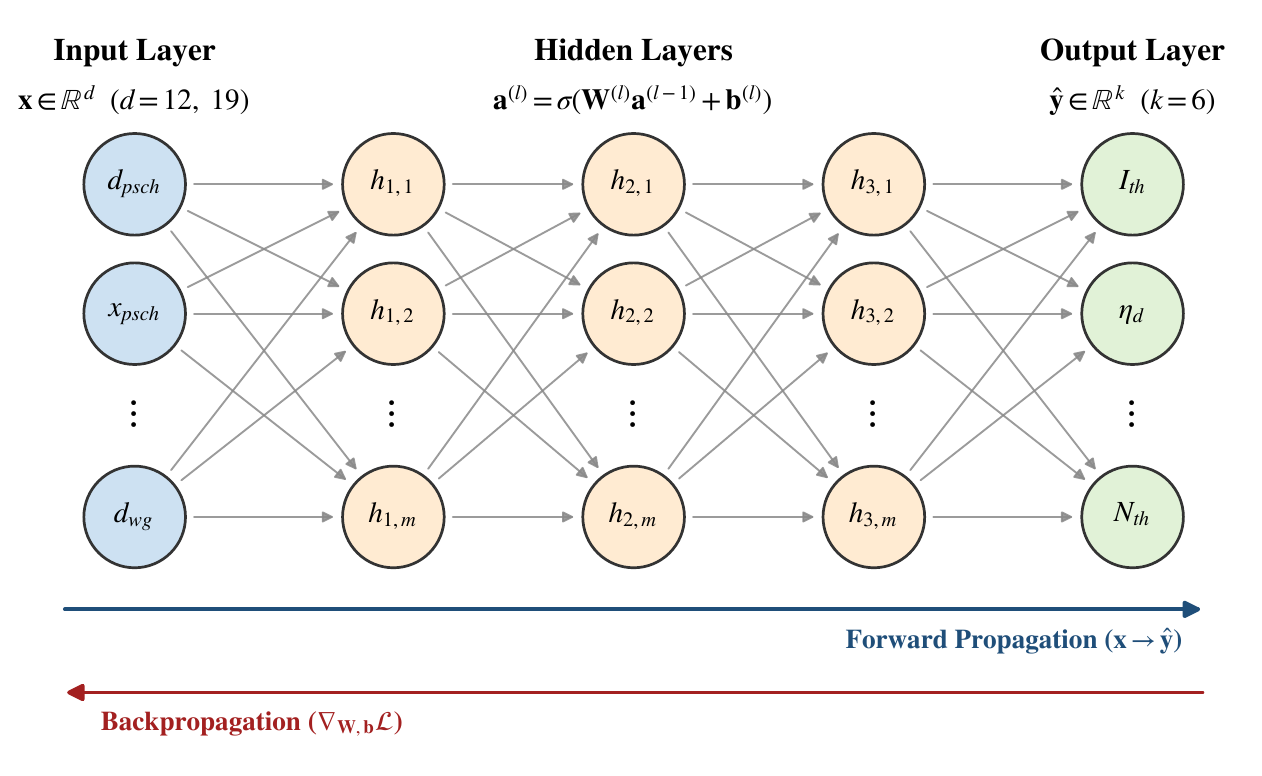}
\caption{Multi-layer perceptron architecture used as the surrogate model.}
\label{fig_NN}
\end{figure}

As illustrated in Fig.~\ref{fig_NN}, the MLP consists of an input layer, three hidden layers, and an output layer. The input vector $\mathbf{x} \in \mathbb{R}^{d}$, representing either the 12- or 19-parameter epitaxial feature space, is propagated forward through the layers. For any given
layer $l$, the activation vector $\mathbf{a}^{(l)}$ is
\begin{equation}
\mathbf{a}^{(l)} = \sigma\!\left( \mathbf{W}^{(l)} \mathbf{a}^{(l-1)}
+ \mathbf{b}^{(l)} \right),
\label{eq:mlp-layer}
\end{equation}
where $\mathbf{W}^{(l)}$ is the weight matrix connecting layer $l-1$ to layer $l$, $\mathbf{b}^{(l)}$ is the bias vector, $\mathbf{a}^{(l-1)}$ is the output of the previous layer (with $\mathbf{a}^{(0)} = \mathbf{x}$), and $\sigma(\cdot)$ is
an activation function such as the rectified linear unit (ReLU) or Swish \cite{ramachandranSearchingActivationFunctions2017}. The activation function introduces the non-linearity that allows the network to approximate non-linear relations between the epitaxial parameters.

Each hidden block in this work applies batch normalization between the linear map and the activation, so that $\mathbf{a}^{(l)} = \sigma(\mathrm{BN}(\mathbf{W}^{(l)}\mathbf{a}^{(l-1)} + \mathbf{b}^{(l)}))$. The input epitaxial parameters in this work span a wide dynamic range, with vastly different physical units. Batch normalization standardizes the pre-activations of each layer across the mini-batch, which stabilizes training over this range. The network inputs and the six targets are likewise standardized to zero mean and unit variance, with the normalization statistics taken from the training split only; predictions are returned to physical units before evaluation.

During the forward pass, the input features pass through the network layers and produce a prediction vector $\hat{\mathbf{y}}$. The model is trained in a loop: a loss function $\mathcal{L}(\mathbf{y}, \hat{\mathbf{y}})$, such as the mean squared error (MSE), quantifies the difference between the predictions and the actual simulation targets. During backpropagation, the partial derivatives of the loss with respect to each weight and bias, $\partial \mathcal{L} / \partial \mathbf{W}$ and $\partial \mathcal{L} / \partial \mathbf{b}$, are computed via the chain rule. An optimization algorithm - Adam, in our case - then updates the model parameters to minimize the total training loss until convergence is reached.

The 4999 data points were split 70/15/15 into training, validation, and test sets and the models are trained with Matlab using the Deep Learning Toolbox. The validation set is used for early stopping and model selection during model training, and all reported metrics are computed from the held-out test set (i.e. samples the model has not seen during the training process).
After training, inference is executed on the test set to generate parity plots comparing the true simulation outputs $y$ against the surrogate predictions $\hat{y}$. Model accuracy is quantified primarily using the coefficient of determination, $R^{2}$, defined as
\begin{equation}
R^{2} = 1 - \frac{\sum_{i=1}^{N} (y_i - \hat{y}_i)^{2}}
                 {\sum_{i=1}^{N} (y_i - \bar{y})^{2}},
\label{eq:r2}
\end{equation}
where $y_i$ is the true simulation value for sample $i$, $\hat{y}_i$ is the corresponding surrogate prediction, $\bar{y}$ is the mean true simulation value across the $N$ test samples, and $R^{2} = 1$ denotes perfect prediction. Parity plots for the models we trained are presented in the next section.

\section{Results}

\subsection{Surrogate Model Performance}

\begin{figure*}[!t]
\centering
\includegraphics[width=6in]{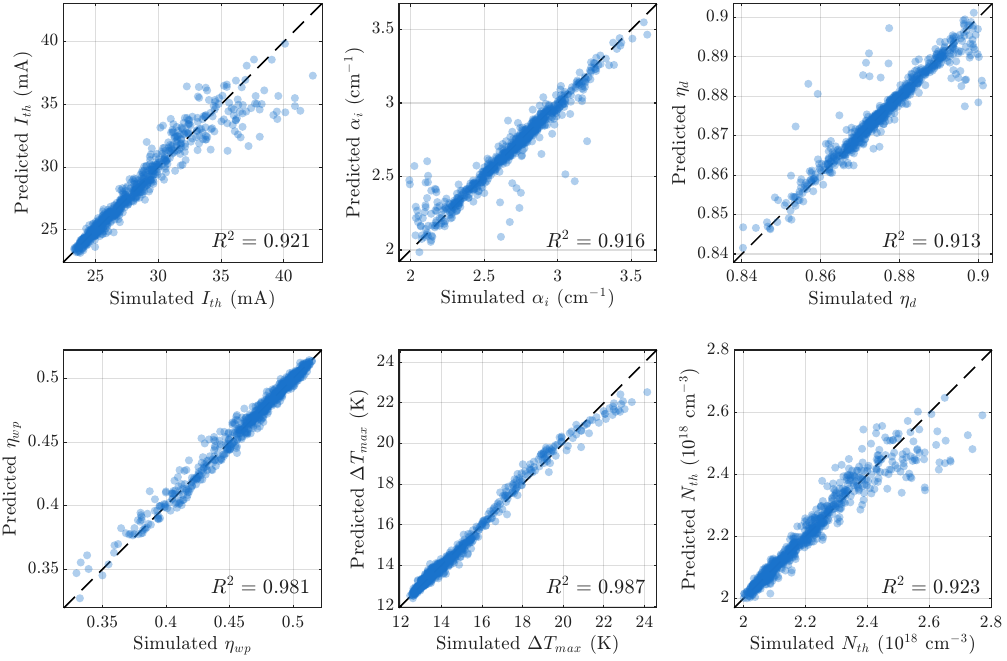}
\caption{Parity plots for the 6 scalar target metrics on the 750 held-out test
samples.}
\label{fig_parity_plots}
\end{figure*}

Fig. \ref{fig_parity_plots} shows the parity plots comparing predicted values to actual values, for the best model we trained, which includes Swish activation and the 7 engineered input features. These plots are generated from the 750-sample held-out test set, which the model did not see during training. $R^2$ is above 0.91 for each of the 6 targets with mean absolute percentage error (MAPE) ranging from 0.19-1.92$\%$, as summarized in Table \ref{tab:metrics}. The range column gives
the span of each target across the test rows:
$R^2 = 1 - (\mathrm{RMSE}/\sigma)^2$.
Note that the MAPE values reported in Table \ref{tab:metrics} are not comparable across targets and should be interpreted with the total range in mind. For example, $\eta_d$ shows the lowest MAPE because the range is small (0.84-0.90) even though its $R^2$ value is the lowest of the 6 targets.
When looking at Fig. \ref{fig_parity_plots} it is also worth noting that some of these 6 targets are closely related to each other. $I_{th}$ and $N_{th}$ are naturally related, particularly for lasers with the same cavity length and width. $\alpha_i$ and $\eta_d$ are also closely related via equation 2.
The model's performance improvement from Swish activation and engineered features is quantified in the next section.

\begin{table}[!t]
\caption{Surrogate model performance on the 750-sample held-out test set
(swish activation, 19 features, 3499 training samples).}
\label{tab:metrics}
\centering
\footnotesize
\setlength{\tabcolsep}{5pt}
\renewcommand{\arraystretch}{1.2}
\begin{tabular}{@{}lcccc@{}}
\toprule
Target & Test-set range & $R^2$ & RMSE & MAPE (\%) \\
\midrule
$I_{th}$ (mA)                        & 23.3--42.3   & 0.9213 & 1.06   & 1.92 \\
$\alpha_i$ (cm$^{-1}$)               & 1.99--3.61   & 0.9161 & 0.084  & 1.65 \\
$\eta_d$                             & 0.840--0.901 & 0.9133 & 0.0033 & 0.19 \\
$\eta_{wpe\,max}$                    & 0.330--0.515 & 0.9809 & 0.0050 & 0.78 \\
$\Delta T_{max}$ (K)              & 12.5--24.1   & 0.9873 & 0.263  & 1.13 \\
$N_{th}$ ($10^{18}$ cm$^{-3}$)       & 2.00--2.77   & 0.9233 & 0.042  & 1.03 \\
\midrule
\textbf{Mean} & -- & \textbf{0.9404} & -- & \textbf{1.12} \\
\bottomrule
\end{tabular}
\end{table}

\subsection{Model Configuration and Data Efficiency}
The parity plots in Fig. \ref{fig_parity_plots} show the best configuration of the model. We now compare the various configurations to quantify the performance improvement from using Swish vs. ReLU and from the 7 engineered input features. To do this, we trained 4 different variations: the base MLP with ReLU activation, first with just the 12 raw input features, then including the engineered features for 19 total inputs; and MLP with Swish activation, with and without the 7 engineered features. We also varied the seed which determines the samples in the 70/15/15 split as well as the weight initialization at the beginning of the training loop. 5 different seeds were used for each of the model configurations.
The results are shown in Table \ref{tab:config} which shows mean $R^2$ and mean MAPE - i.e. averaged across the 6 targets - for the 4 different cases. All results shown in the table were for models trained on the full 3499-sample training set, and evaluated on the 750 held-out test samples. The error values in $R^2$ and MAPE come from the 5 different seeds. 
It is clear that both Swish activation and the engineered features lead to notable performance improvements. 
The best performing model uses all 19 features with Swish activation, which raises
mean $R^2$ from $0.879 \pm 0.011$ to $0.938 \pm 0.003$ and lowers mean MAPE from $1.78 \pm 0.08 \%$ to $1.17 \pm 0.09 \%$.
Expressed as error rather than fit, the unexplained variance $1-R^2$ falls from
0.118 to 0.060, a 49\% reduction, and the mean percentage error falls by 36\%.

\begin{table}[!t]
\caption{Effect of Activation Function and Engineered Features.}
\label{tab:config}
\centering
\footnotesize
\setlength{\tabcolsep}{5pt}
\renewcommand{\arraystretch}{1.2}
\begin{tabular}{@{}llcc@{}}
\toprule
Activation & Features & Mean $R^2$ & Mean MAPE (\%) \\
\midrule
ReLU  & 12 raw       & $0.8791 \pm 0.0111$ & $1.78 \pm 0.08$ \\
ReLU  & 19 with eng. & $0.9236 \pm 0.0093$ & $1.42 \pm 0.11$ \\
Swish & 12 raw       & $0.9138 \pm 0.0091$ & $1.37 \pm 0.05$ \\
\textbf{Swish} & \textbf{19 with eng.} & $\mathbf{0.9384 \pm 0.0031}$ & $\mathbf{1.17 \pm 0.09}$ \\
\bottomrule
\end{tabular}
\end{table}

\begin{figure}[!t]
\centering
\includegraphics[width=3in]{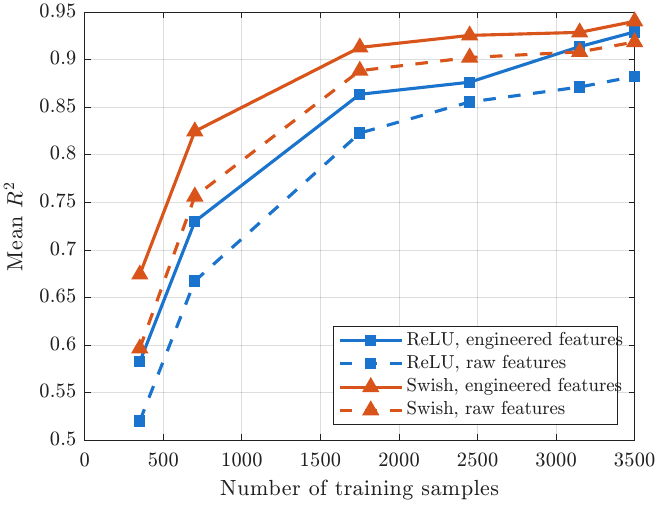}
\caption{$R^2$ versus number of training samples for the 4 different model configurations.}
\label{fig_data_eff}
\end{figure}

We also compared how model performance depends on the size of the data set. Fig. \ref{fig_data_eff} compares the same 4 models by plotting mean $R^2$ vs. the number of training samples, from 350 to 3499 (the full training split). The results from Table \ref{tab:config} are consistent as the size of the data set varies, with the 19-input Swish model demonstrating the best performance. In all cases, performance drops rapidly when the number of samples is reduced below 1750. Encouragingly, the 19-input Swish configuration still yields mean $R^2$ of 0.91 when the data set is cut in half from 3499 to 1750 samples. It is worth noting that this is the \textit{mean} value, and it is dominated somewhat by targets that are easier to predict such as $\Delta T_{max}$. However, in this case, even the least accurate metric, $\eta_d$, still yields $R^2 = 0.87$. It should also be noted that the relative improvement from Swish and engineered features becomes more pronounced as the number of samples is reduced with a delta of 0.09 between the best and worst models when trained on 1750 samples, vs. a delta of 0.06 when trained on the full data set. The difference is even more pronounced at the smallest sample size of 350 samples, where the delta between best and worst is 0.15.

\section{Epitaxy Optimization with Surrogate Model Acceleration}
A single inference with the surrogate model takes approximately 0.015 ms on a laptop with Intel Core i5-13600H CPU. Compared to a Harold simulation of a similar design, which takes approximately 20 seconds on the same laptop, this represents a $\sim10^6\times$ speed up. In this section we present an approach to leverage this speed-up for accelerated laser design.

\subsection{Pareto Search and Validation with Harold}
\begin{figure}[!t]
\centering
\includegraphics[width=3in]{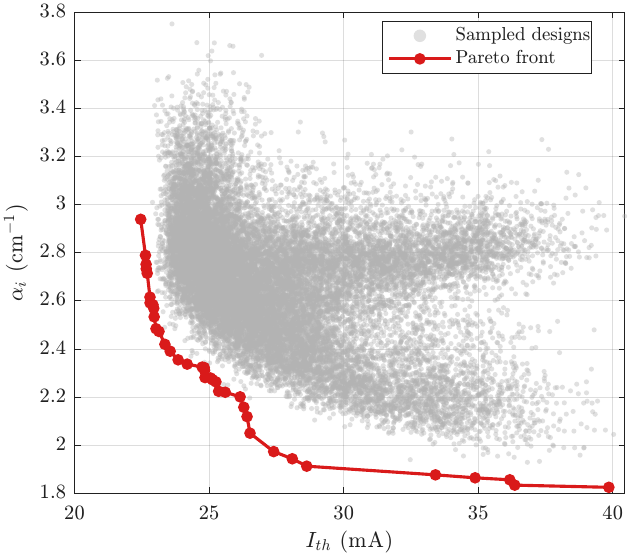}
\caption{Densely sampled parameter space to search for designs that simultaneously minimize $I_{th}$ and $\alpha_i$. The red curve in this plot represents the Pareto front.}
\label{fig_pareto}
\end{figure}

A simple ``brute-force" approach to using the surrogate is to sample the input parameter space much more densely, and search for designs that optimize for a specific metric, or metrics. Here, as an example, we have performed a Pareto search where we sample the input parameter space with 1 million designs using LHS, predict the output for each design and search the results for designs that simultaneously minimize both $I_{th}$ and $\alpha_i$. The result is a Pareto frontier of designs that minimize these two values. Fig. \ref{fig_pareto} plots $\alpha_i$ vs. $I_{th}$ where each point on the plot represents a unique epitaxy design. The red curve highlights the Pareto frontier which represents all designs where improvements in either $\alpha_i$ or $I_{th}$ will necessarily degrade the other. For the 1 million designs sampled here, this yielded 37 designs along the frontier. Inference for these 1 million designs using the surrogate model took approximately 2 minutes (including search and sort times) on a laptop, vs. 231 days which would be required to simulate all these designs in Harold. 

\begin{figure}[!t]
\centering
\includegraphics[width=3in]{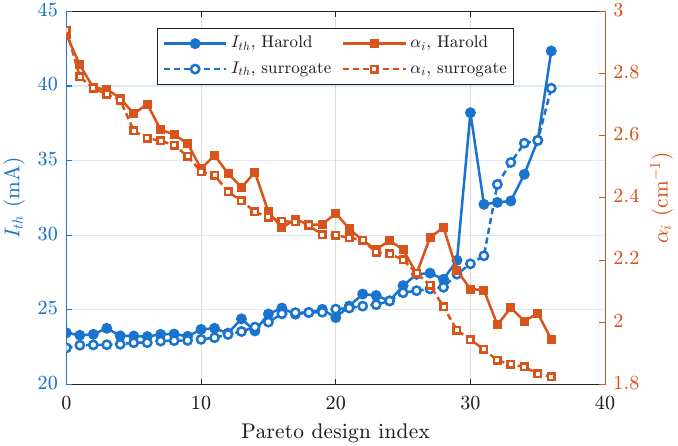}
\caption{Simulated and surrogate predicted $I_{th}$ and $\alpha_i$ for the 37 Pareto optimized designs.}
\label{fig_validate_pareto}
\end{figure}

The 37 designs found by the Pareto search were then simulated with Harold to compare the surrogate model's inference to simulation. These results are plotted in Fig. \ref{fig_validate_pareto} which plots simulated and inferred $I_{th}$ and $\alpha_i$ vs. sample index. It can be seen here that simulation and prediction track very closely, although with some outliers.
The MAPE for each of the 6 targets, averaged across the 37 samples, is shown in Table \ref{tab:pareto}, along with percent bias (indicating whether the model tends to over or under predict), and the maximum error percentage. The error is signed, $(\hat{y}-y)/y$, so a negative bias indicates the surrogate under-predicts that value.
The average MAPE between inference and simulation is 3.39\% for $I_{th}$ and 3.19\% for $\alpha_i$, compared to 1.92\% and 1.65\% for $I_{th}$ and $\alpha_i$, respectively, on the held-out test in Table \ref{tab:metrics}. The remaining 4 targets show even lower MAPE. 
Some degradation is expected since the Pareto front is drawn from the extremes of the objective space, so these 37 designs likely sit close to the boundary of the sampled parameter box, where the surrogate has the least surrounding training data to interpolate from. The agreement in Fig. \ref{fig_validate_pareto} is consequently a harder test of the model than the held-out test metrics, yet MAPE still remains at reasonable level for most designs.

\begin{table}[!t]
\caption{Surrogate Accuracy on the 37 Pareto-Optimal Designs, Re-Simulated in Harold.}
\label{tab:pareto}
\centering
\footnotesize
\setlength{\tabcolsep}{5pt}
\renewcommand{\arraystretch}{1.2}
\begin{tabular}{@{}lccc@{}}
\toprule
Target & MAPE (\%) & Bias (\%) & Max $|$err$|$ (\%) \\
\midrule
$I_{th}$         & 3.39 & $-2.22$            & 26.50 \\
$\alpha_i$       & 3.19 & $-3.11$            & 11.06 \\
$\eta_d$         & 0.29 & $\phantom{-}0.27$  & 1.21 \\
$\eta_{wpe,~max}$      & 1.31 & $\phantom{-}0.75$  & 13.98 \\
$\Delta T_{max}$ & 1.23 & $\phantom{-}0.15$  &  4.92 \\
$N_{th}$         & 1.64 & $-1.04$            & 16.51 \\
\bottomrule
\end{tabular}
\end{table}

Note that sample index is an arbitrary numerical label that tells nothing about the actual epitaxy design parameters for that design (the points in Fig. \ref{fig_validate_pareto} are simply ordered by $I_{th}$ and $\alpha_i$). When looking at the epitaxial layer parameters, however, the Pareto optimized designs do indeed make intuitive physical sense. For example, $I_{th}$ generally increases as the waveguide thickness increases, and $\Delta T_{max}$ increases with reduced doping in the cladding layers.

\subsection{Domain of Validity}

\begin{figure}[!t]
\centering
\includegraphics[width=3in]{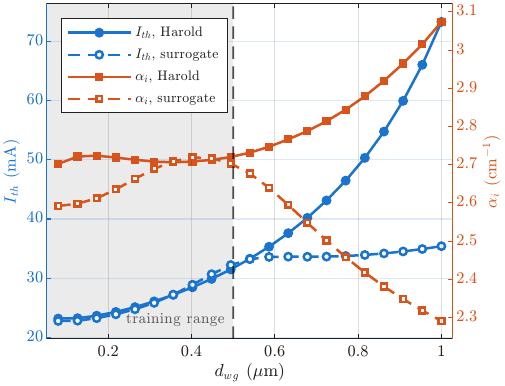}
\caption{Simulation vs. surrogate prediction for the low-$I_{th}$ design from Fig. \ref{fig_validate_pareto} with waveguide thickness swept beyond the range of the training data set to show divergence.}
\label{fig_domain}
\end{figure}

A natural question that arises at this point is, ``how well does the model perform on designs outside the training parameter space?" To test this, we chose a fixed design as a starting point, swept waveguide thickness beyond the range of the training data, and plotted simulated and predicted $I_{th}$ and $\alpha_i$. 
This is shown in Fig. \ref{fig_domain}. In this example, the starting point is the lowest $I_{th}$ design from Fig. \ref{fig_validate_pareto} - the design found in the Pareto search for minimizing threshold current. The initial waveguide thickness, $d_{wg}$, is 80 nm and was increased out to 1 $\mu m$ with all other layers held constant. The parameter space for the training data set had a maximum waveguide thickness of 500 nm and it can clearly be seen from Fig. \ref{fig_domain} that the Harold simulation and surrogate prediction begin to diverge when the waveguide thickness is made larger than the range included in training.  
It should be noted that the rate at which the surrogate model diverges from simulation truth likely depends on which parameters are swept beyond the training range. This was not thoroughly explored here, however, these results do confirm that, unsurprisingly, the surrogate's prediction is not trustworthy outside the range of the training data. Most notably, the surrogate fails to even predict the correct direction in the trend of $\alpha_i$ as soon as $d_{wg}$ extends beyond 500 nm.

\section{Discussion and Conclusion}
In this work we have demonstrated, for the first time to our knowledge, a surrogate model that predicts edge emitting FP QW laser performance from epitaxial layer properties. The surrogate is based on a simple MLP architecture and achieves $R^2$ greater than 0.91 for all 6 of the predicted targets, with a mean value of 0.94 averaged across the 6 targets. 
We also showed that the inclusion of engineered input features and the use of Swish activation during training raise the mean $R^2$ from 0.88 to 0.94,
removing 49\% of the unexplained variance and 36\% of the mean percentage error without additional simulations.

In considering the limitations of our approach, it should be noted that the surrogate model demonstrated here is only valid for the epitaxial parameter space in Table \ref{tab:epitaxy} which has no variation in the active region, and a fixed laser geometry (600 $\mu m$ X 25 $\mu m$). However, the parameter space does cover a relatively wide range of potential designs, from very thin, tightly confined waveguides (e.g. 20 nm total waveguide thickness, surrounded by SCH layers with 35\% Al fraction) to fairly thick and dilute waveguides (500 nm thick, surrounded by low-confinement SCH layers with 10\% Al fraction). 
Furthermore, the surrogate reaches this accuracy from only 3499 training samples. In a 12-dimensional design space, a two-level full factorial over the same parameters would be $2^{12} = 4096$ samples. So, the 3499-sample training set correspond to fewer than two levels per axis, yet still appears to capture the underlying physics reasonably well.
For epitaxial designs with fewer variables of interest, the number of required training simulations may be even less. This approach enables sparse sampling and quick model training, followed by rapid inference over many designs, and final validation on a curated subset. The data efficiency plots in Fig. \ref{fig_data_eff} are promising as they indicate that one only needs 1750 samples to maintain average $R^2$ greater than 90\%, and the curves rise slowly with increasing data after that. 
Another potential improvement to the model could be the addition of physics informed loss functions during training. With this approach, rather than the simple MSE loss function we used here, the training algorithm could utilize relevant physical equations and penalize the model for predicting values that violate these formulae. We plan to explore this approach in future work. 

Regarding accuracy, this is a surrogate model of Harold, and it inherits any assumptions or approximations made in the simulation. So, comparing surrogate model predictions to real device measurements would yield the inherent error between simulated and real devices (e.g. due to fabrication tolerances), plus the additional error between the surrogate model and Harold, which is what we have quantified in this paper. We have not yet checked accuracy against real device measurements, however, we plan to make this comparison in future work. 

In conclusion, we have demonstrated that a simple neural network can be trained as a surrogate model to predict laser performance with a relatively small data set in a 12-dimensional epitaxial design space. Inference with the surrogate is 6 orders of magnitude faster than running a simulation, and enables rapid exploration of the design space for multi-variable optimization. We expect this approach to generalize beyond our specific design, and it provides a facile technique for optimizing semiconductor laser epitaxy designs with sparse sampling. 

\section{Code and Data Availability}
The MATLAB code used to train the neural network is available at https://doi.org/10.5281/zenodo.22897556, \cite{verrinder2026code}. This record also contains scripts for the Pareto search and data for reproducing the plots in Figs. \ref{fig_validate_pareto} and \ref{fig_domain}. The full training dataset was generated using third-party licensed software and neither this nor the trained model are released. However, the code is distributed with synthetic example data matching the structure of the training set to test the workflow.

\section*{Acknowledgments}
The authors acknowledge the use of Claude from Anthropic and Gemini from Google to assist with writing the code used for automating the simulation loop and training the models in Sections II, III, and IV. All code was reviewed and validated by the authors, who take full responsibility for the results presented herein. All text and figures in the manuscript were prepared entirely by the authors.

\bibliographystyle{IEEEtran}
\bibliography{references}

\newpage

\section{Biography Section}
 
\vspace{11pt}

\begin{IEEEbiography}[{\includegraphics[width=1in,height=1.25in,clip,keepaspectratio]{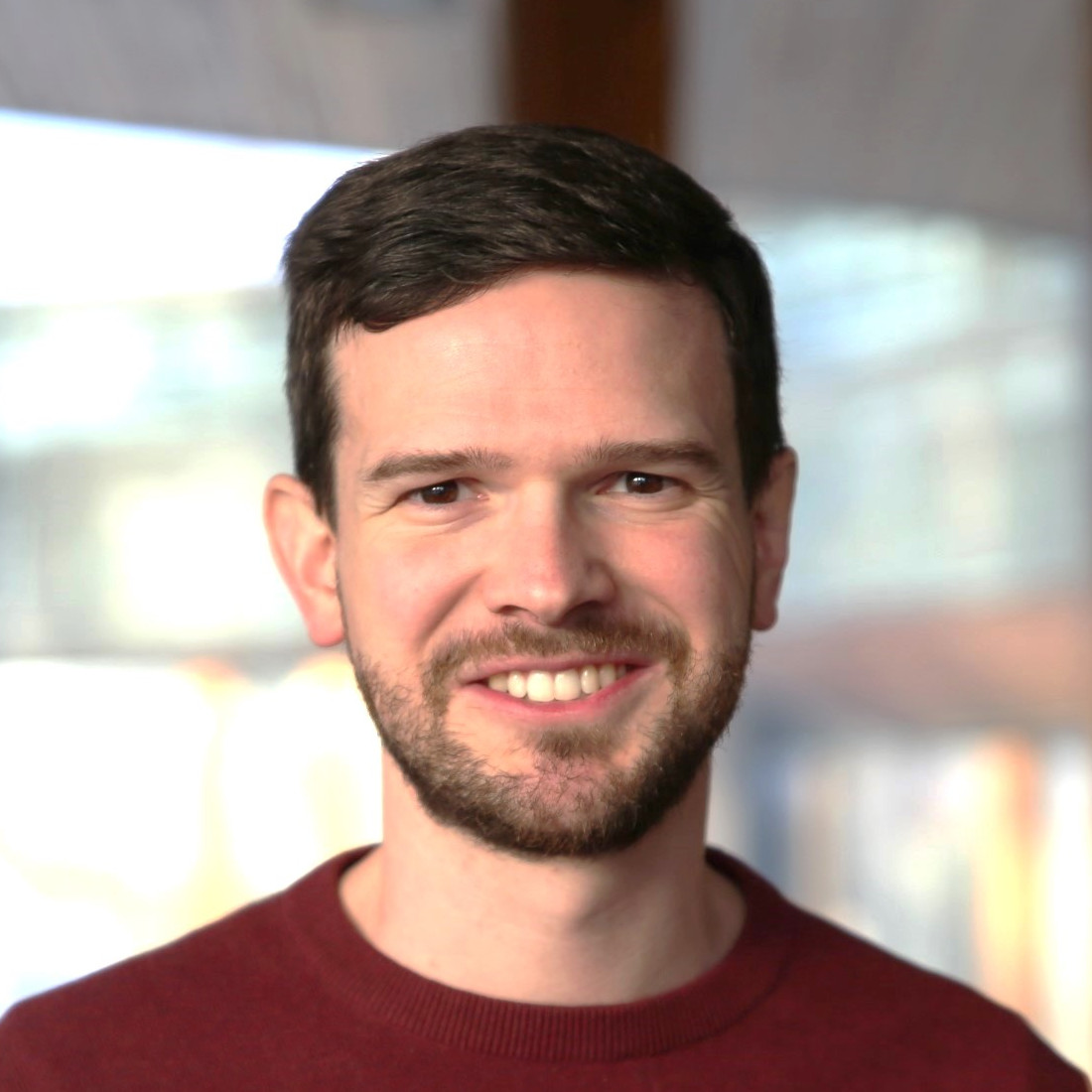}}]{Paul Verrinder}
received the B.S. degree in electrical engineering from California State University, Fresno, Fresno, CA, USA and the M.S. and Ph.D. degrees from the University of California Santa Barbara (UCSB), Santa Barbara, CA, USA in 2020 and 2023, respectively. From 2023 to 2024 he was a postdoctoral researcher at Aalto University and in 2024 he joined the Department of Electronics and Nanoengineering at Aalto University, Espoo, Finland as an Assistant Professor.
\end{IEEEbiography}

\vspace{11pt}

\begin{IEEEbiography}[{\includegraphics[width=1in,height=1.25in,clip,keepaspectratio]{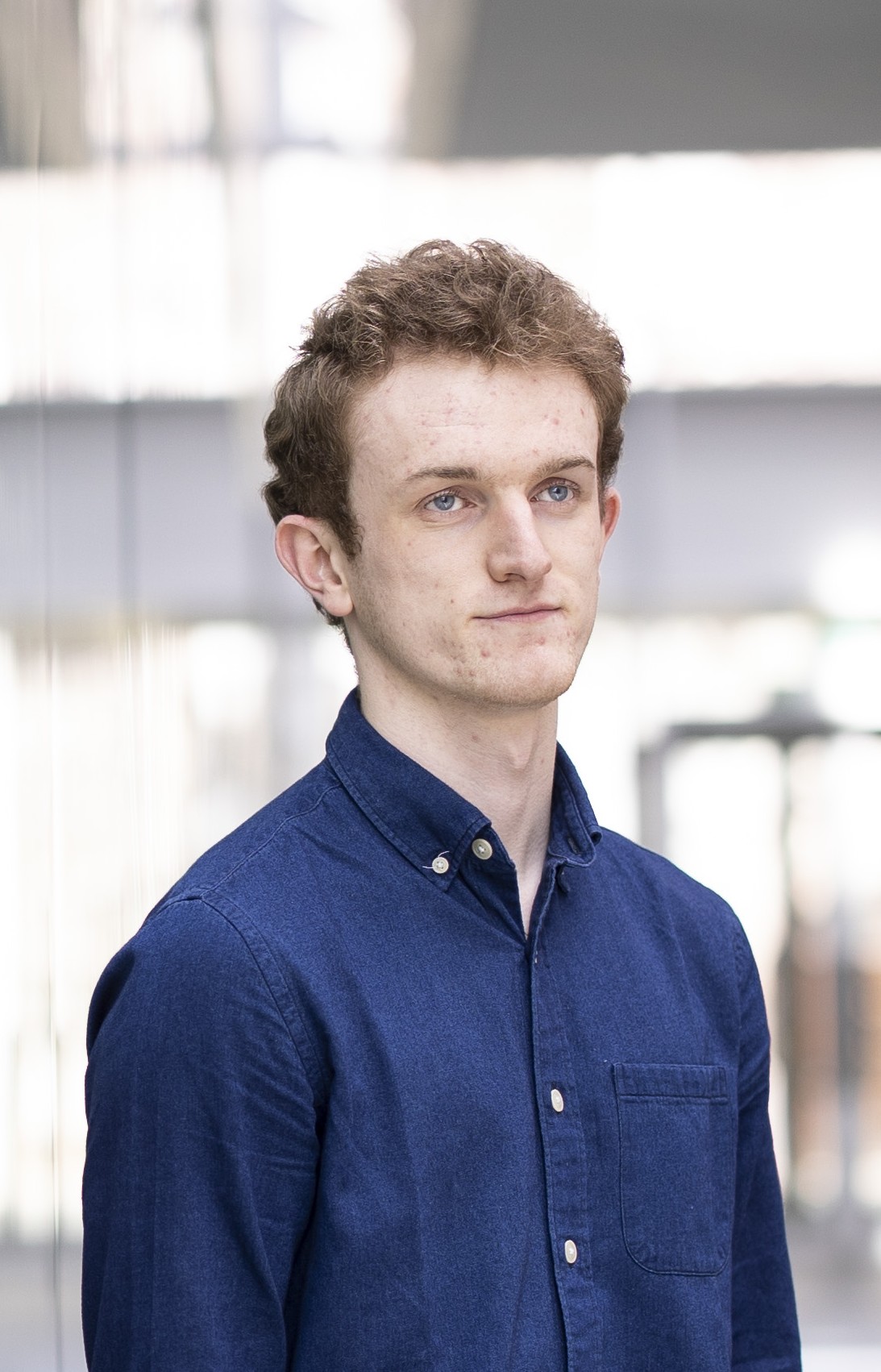}}]{Jack Markham}
received the B.Sc. degree in Applied Physics from the University of Limerick, Limerick City, Ireland in 2025. He is an Erasmus Mundus scholar currently pursuing a M.Sc. in the iPSRS degree program.
\end{IEEEbiography}

\vfill

\end{document}